\documentclass[journal]{IEEEtran}
\usepackage{cite}

\usepackage{graphicx}
\DeclareGraphicsExtensions{.pdf,.jpeg,.png,.jpg, .svg}

\usepackage{amssymb}
\usepackage[cmex10]{amsmath}
\usepackage{array}

\usepackage[font = small]{caption}
\usepackage{color}
\usepackage{commath}
\usepackage{amssymb}
\usepackage{bm}
\usepackage{amsmath}

\usepackage{mathtools}

\newcommand{\etal}{\textit{et al}. }
\newcommand{\ie}{\textit{i}.\textit{e}. }

\usepackage[utf8]{inputenc}
\usepackage{algorithm}

\usepackage[]{algpseudocode}

\algnewcommand\algorithmicinput{\textbf{Input:}}
\algnewcommand\Input{\item[\algorithmicinput]}
\algnewcommand\algorithmicoutput{\textbf{Output:}}
\algnewcommand\Output{\item[\algorithmicoutput]}
\algnewcommand\algorithmichline{}
\algnewcommand\Hline{\item[\algorithmichline]}
\makeatletter
    \newcommand*{\algrule}[1][\algorithmicindent]{\makebox[#1][l]{\hspace*{.5em}\thealgruleextra\vrule height \thealgruleheight depth \thealgruledepth}}%
\newcommand*{\thealgruleextra}{}
\newcommand*{\thealgruleheight}{.85\baselineskip}
\newcommand*{\thealgruledepth}{.25\baselineskip}

\newcount\ALG@printindent@tempcnta
\def\ALG@printindent{%
    \ifnum \theALG@nested>0
        \ifx\ALG@text\ALG@x@notext
        \else
            \unskip
            \addvspace{-1pt}
            \ALG@printindent@tempcnta=1
            \loop
                \algrule[\csname ALG@ind@\the\ALG@printindent@tempcnta\endcsname]%
                \advance \ALG@printindent@tempcnta 1
            \ifnum \ALG@printindent@tempcnta<\numexpr\theALG@nested+1\relax
            \repeat
        \fi
    \fi
    }%
\usepackage{etoolbox}
\patchcmd{\ALG@doentity}{\noindent\hskip\ALG@tlm}{\ALG@printindent}{}{\errmessage{failed to patch}}
\makeatother

\newbox\statebox
\newcommand{\myState}[1]{%
    \setbox\statebox=\vbox{#1}%
    \edef\thealgruleheight{\dimexpr \the\ht\statebox+1pt\relax}%
    \edef\thealgruledepth{\dimexpr \the\dp\statebox+1pt\relax}%
    \ifdim\thealgruleheight<.75\baselineskip
        \def\thealgruleheight{\dimexpr .75\baselineskip+1pt\relax}%
    \fi
    \ifdim\thealgruledepth<.25\baselineskip
        \def\thealgruledepth{\dimexpr .25\baselineskip+1pt\relax}%
    \fi
    \State #1%
    \def\thealgruleheight{\dimexpr .75\baselineskip+1pt\relax}%
    \def\thealgruledepth{\dimexpr .25\baselineskip+1pt\relax}%
}

\algdef{SE}[DOWHILE]{Do}{doWhile}{\algorithmicdo}[1]{\algorithmicwhile\ #1}%

\begin{document}
\title{Resource Allocation for Elastic Optical Networks using Geometric Optimization}
\author{Mohammad~Hadi,~\IEEEmembership{Member,~IEEE,}
        and~Mohammad~Reza~Pakravan,~\IEEEmembership{Member,~IEEE}
\thanks{Mohammad Hadi is a PhD student at Department of Electrical Engineering, Sharif University of Technology, e-mail: mhadi@ee.sharif.edu}
\thanks{Mohammad Reza Pakravan is with Department of Electrical Engineering, Sharif University of Technology as an associate professor, e-mail: pakravan@sharif.edu}}

\maketitle

\begin{abstract}
\boldmath
Resource allocation with quality of service constraints is one of the most challenging problems in elastic optical networks which is normally formulated as a mixed-integer nonlinear optimization program. In this paper, we focus on novel properties of geometric optimization and provide a heuristic approach for resource allocation which is very faster than its mixed-integer nonlinear counterpart. Our heuristic consists of two main parts for routing/traffic ordering and power/spectrum assignment. It aims at minimization of transmitted optical power and spectrum usage constrained to quality of service and physical requirements. We consider three routing/traffic ordering procedures and compare them in terms of total transmitted optical power, total received noise power and total nonlinear interference including self- and cross-channel interferences. We propose a posynomial expression for optical signal to noise ratio in which fiber nonlinearities and spontaneous emission noise have been addressed. We also propose posynomial expressions that relate modulation spectral efficiency to its corresponding minimum required optical signal to noise ratio. We then use the posynomial expressions to develop six geometric formulations for power/spectrum assignment part of the heuristic which are different in run time, complexity and accuracy. Simulation results demonstrate that the proposed solution has a very good accuracy and much lower computational complexity in comparison with mixed-integer nonlinear formulation. As example for European Cost239 optical network with $46$ transmit transponders, the geometric formulations can be more than $59$ times faster than its mixed-integer nonlinear counterpart. Numerical results also reveal that in long-haul elastic optical networks, considering the product of the number of common fiber spans and the transmission bit rate is a better goal function for routing/traffic ordering sub-problem.
\end{abstract}

\begin{IEEEkeywords}
Posynomial expression, Geometric optimization, Elastic optical networks, Resource allocation, Quality of service.
\end{IEEEkeywords}

\IEEEpeerreviewmaketitle

\section{Introduction}\label{sec_I}
\IEEEPARstart{T}{raditional} fixed grid optical networks are not adaptive and therefore, they can not efficiently use systems resources such as power and spectrum according to the conditions of diverse heterogeneous traffic demands. To meet the requirements of constantly increasing and time-variable demands of data traffic, Elastic Optical Networks (EON) are used to adaptively assign routes, bandwidths, modulation levels and transmitted optical powers. Flexible assignment of such a wide range of system resources which are inter-dependent is a complex optimization problem and is conventionally referred to as the Routing and Spectrum Assignment (RSA). Providing fast and efficient RSA algorithms is an important topic of research \cite{chatterjee2015routing, abkenar2017study, sambo2015routing}.

Several variants of optimization formulations and algorithms have been proposed for RSA in EON where system variables are optimally selected such that a cost function (such as spectrum usage or power consumption) is minimized,  physical constraints (such as spectrum continuity, spectrum contiguity and spectrum non-overlapping) are satisfied and pre-defined levels of QoS (which is usually translated to equivalent levels of Optical Signal to Noise Ratio (OSNR)) are guaranteed. Some of the proposed solutions consider the optical fiber as an ideal channel and do not consider the QoS requirements in their optimization analysis \cite{fallahpour2015energy, hadi2016improved, archambault2016routing}. This is an oversimplification of the real world issues and does not yield optimized practical results. Fiber is a non-ideal channel and optical fiber communication requires detailed attention to the channel effects such as Amplified Spontaneous Emission (ASE) noise and NonLinear Interferences (NLI). Optical transmit power is a very important parameters in ensuring proper QoS level. It affects the nonlinear behavior of the fiber, transmission distance, modulation level and spectrum usage. Although some researchers have considered QoS, they do not consider the transmit power as an optimization variable which results in inefficient power allocation because of ignoring one degree of freedom in the optimization \cite{beyranvand2013quality, khodakarami2016quality, khodakarami2014flexible, yan2015link, zhao2015nonlinear}. It has been shown that by optimizing the transmission power per connection, the spectrum usage can be reduced by around $20\%$ \cite{yan2017joint}. Power optimized Routing and Wavelength Assignment (RWA) in nonlinear Wavelength Division Multiplex (WDM) networks has been studied in \cite{roberts2016convex, ives2014physical, ives2014adapting, ives2015routing}. Although these algorithms improve network throughput, they are specifically designed for WDM networks and cannot be applied to EONs with variable spectrum bandwidths and carrier frequency locations. Few research works take into account the interaction between flexible resources, ASE noise and NLIs to optimally allocate available resources, especially transmission power, while satisfying QoS requirements \cite{yan2015resource, yan2017joint}. Yan \etal have studied resource allocation for EONs with nonlinear channel model \cite{yan2015resource}. Their work decomposes resource allocation problem into two sub-problems of 1) Routing/Traffic Ordering (RTO) and 2) Power/Spectrum Assignment (PSA) and provides a Mixed-Integer Nonlinear Program (MINLP) formulation for PSA. Unfortunately, MINLP is an NP-hard problem and such formulation is not a suitable choice for practical resource provisioning in EONs. To overcome the complexity of MINLP formulation, a Mixed Integer Linear Program (MILP) formulation has been proposed in \cite{yan2017joint} where a piecewise linear approximation for nonlinear terms of the conventional MINLP formulation is used. Although their MILP optimization problem can be more efficiently solved but its relative error may be high especially when the number of linear functions in piecewise linear approximation is low. Some researchers have also evaluated the impact of different RTO scenarios on spectrum usage \cite{yan2015resource, zhao2015nonlinear} but to the best of our knowledge, no one has provided a comprehensive investigation of the effect of RTO on NLIs and ASE noise. 

A Geometric Program (GP) is a mathematical optimization problem characterized by simple and generalized posynomial objective and constraint functions that have a special form. A geometric program is simply converted to a convex optimization problem and therefore, large-scale GPs are efficiently and reliably solved using advanced convex optimizer software packages and algorithms \cite{boyd2007tutorial, boyd2004convex}. GP is successfully used in QoS-aware communication optimization problems \cite{chiang2005geometric}. Following the approach of \cite{yan2015resource} and many other papers, we have decomposed the resource allocation problem into two interconnected optimization problems, \ie RTO and PSA. In the first problem, we optimally assign routing and traffic orders and in the second problem, we assign spectrum and power parameters. We consider NLI and ASE noise in our formulation and provide posynomial expressions to describe their impact on QoS. We also use few generalized posynomial expressions to relate the value of modulation spectral efficiency to its corresponding minimum required OSNR. Then, we use the proposed posynomial expressions to formulate six GPs for PSA sub-problem which have different accuracy, efficiency and run time. We also consider three RTO procedures which are distinguished by their cost functions. The proposed heuristic provides accurate solutions with a considerable lower complexity in comparison with the conventional MINLP approach. As an example, we have used European Cost239 network topology and evaluated the complexity and error of the proposed algorithms. Results show that our proposed GP formulations for PSA can be more than one order of magnitude faster and have negligible error compared to the MINLP. We have also used simulations to compare the three proposed RTO procedures and show the best candidate.

The rest of the paper is organized as follows. System model is introduced in Section \ref{Sec_II}. We propose our two-stage heuristic for resource allocation in Section \ref{Sec_III}. RTO procedures are discussed in Section \ref{Sec_IV} while we develop our GP formulations for PSA sub-problem in \ref{Sec_V}. Simulation results are included in \ref{Sec_VI}. Finally, we conclude the paper in Section \ref{Sec_VII}.

\textbf{Notation}. Optimization variables are shown in lower case letters. Bold letters are used to denote vectors and sets. We use calligraphic font to show constants. Functions are also shown by capital letters and their arguments are given by lower indexes. Optical fiber characterizing parameters are shown by thier typical notations.

\section{System Model}\label{Sec_II}
Consider a coherent optical communication network characterized by topology graph $G(\mathbf{V}, \mathbf{L})$ where $\mathbf{V}$ and $\mathbf{L}$ are the sets of optical nodes and directional optical fiber links, respectively. The optical fiber bandwidth $\mathcal{B}$ is assumed to be gridless. $\mathbf{Q}$ is the set of connection requests and $\mathbf{Q}_l, l \in \mathbf{L}$ shows the set of connection requests that share optical fiber $l$ on their allocated routes. Each connection request $q \in \mathbf{Q}$ is given a contiguous spectrum bandwidth $\Delta_q$ around carrier frequency $\omega_q$. To facilitate optical switching and remove the high cost of spectrum conversion \cite{spectrum2017hadi}, we assume that the assigned spectrum bandwidth to connection request $q$ is continuous over its routed path. $q$-th connection request passes $\mathcal{N}_q$ fiber spans along its routed path and has $\mathcal{N}_{q,i}$ shared fiber spans with connection request $i\neq q$. There are pre-defined modulation formats where each format has spectral efficiency $c$ and requires minimum OSNR $\Theta(c)$ to get a pre-Forward-Error-Correction (FEC) Bit-Error-Rate (BER) value of $4\times 10^{-3}$, as shown in Tab. \ref{tab:snr_spc} \cite{yan2015resource}. The optical transponder of connection request $q$ is given a modulation format with spectral efficiency $c_q$ and fills its assigned optical bandwidth $\Delta_q$ with optical power $p_q$. Transponders have maximum information bit rate $\mathcal{C}$ and fill both of the polarizations with the same power. Assuming Nyquist spectrum shaping, we clearly have $\Delta_q = \frac{\mathcal{R}_q}{c_q}$ where $\mathcal{R}_q$ is required traffic volume. According to filtering and switching constraints in optical nodes, we consider a guard band $\mathcal{G}$ between any two adjacent connection requests on each link. 

\begin{table}[t!]
\centering
\caption{Available modulation formats along with their corresponding spectral efficiency $c$ and minimum required OSNR $\Theta(c)$ to achieve a pre-FEC BER values of $4 \times 10^{-3}$.}\label{tab:snr_spc}
\begin{tabular}{ccc}
\hline
\hline
Modulation Format & Spectral Efficiency $c$ & Minimum Required OSNR $\Theta(c)$\\
\hline
PM-BPSK & 2 & 3.52\\
PM-QPSK & 4 & 7.03\\
PM-8QAM & 6 & 17.59\\
PM-16QAM & 8 & 32.60\\
PM-32QAM & 10 & 64.91\\
PM-64QAM & 12 & 127.51\\
\hline
\hline
\end{tabular}
\end{table}

\section{RTO/PSA Problem}\label{Sec_III}
Solving resource allocation problem, the values of system model variables are determined such that a mixed cost function of transmission optical power and spectrum usage is minimized, physical constrains are satisfied and desired levels of OSNR are guaranteed. In general, such a problem is very hard to solve in reasonable time \cite{yan2015resource}. Therefore, we propose the two-stage heuristic Alg. \ref{alg:rsa} in which the complex resource allocation problem is  decomposed into two sub-problems: 1) RTO, where the routing and the ordering of connection requests on each link are defined, and 2) PSA, where resources such as optical power, spectrum width and carrier frequencies are allocated \cite{yan2015resource} . Usually the search for a near optimal solution involves iterations between these two sub-problems. To save this iteration time, it is of great interest to hold the run time of each sub-problem at its minimum value. In this work, we mainly focus on the second sub-problem which is the most time-consuming one and formulate it as a GP problem to benefit from fast convex optimization algorithms. 

In the first stage of Alg. \ref{alg:rsa}, connection request volumes are partitioned such that no traffic volume is greater than transponder information bit rate $\mathcal{C}$. Thus, a high volume connection request is replaced by multiple connection requests that have the same source and destination as the original request but their volumes are no greater than $\mathcal{C}$. At the next step, RTO procedures should be applied. We consider three RTO methods, named Shortest Path Routing (SPR), Shortest Common Path Routing (SCPR) and Shortest Common Path-Rate Routing (SCPRR), which have different cost function and performance. At the next stage of Alg. \ref{alg:rsa}, the problem of PSA is addressed. We propose six Mixed-Integer GP (MIGP) formulations for PSA which are referred to as Geometric PSA (GPSA).  One of the proposed GPSA formulations is selected for the second stage of Alg. \ref{alg:rsa} and its continuous relaxed version is solved in a local loop \cite{boyd2007tutorial}. At each epoch, the relaxed GPSA formulation is optimized and obtained values for relaxed integer variables (here, $c_q$'s) are rounded by precision $\mathcal{J}$. If none of the rounded variables is valid (based on column $c$ in Tab. \ref{tab:snr_spc}), the precision is increased by $\mathcal{J}$ and the rounding is again applied. This process continues to find at least one valid element for relaxed integer variables. Then, we fix the acceptable rounded values and solve the relaxed continuous GPSA again. The local loop continues until all the integer variables have valid and acceptable values. Note that the heuristic needs at most $\abs{\mathbf{Q}}$ iterations (number of integer variables) to evaluate the GPSA formulation before it converges to a solution and in most cases, the number of iterations is less than this maximum. With each iteration of the local loop since some of the variables are fixed, there are fewer variables to be optimized and therefore, the loop runs faster.

In the coming section, we provide a detailed study of the RTO procedures, \ie SPR, SCPR and SCPRR. Then, we focus on PSA and  show how it can be formulated as a GP. 

\begin{algorithm}[t!]
\small
\caption{GPSA Heuristic Algorithm}\label{alg:rsa}
\begin{algorithmic}[1]
\Input{network topology, traffic matrix, link parameters, noise and nonlinearity parameters, transponder parameters, rounding precision}
\Output{routes, frequency spectrums, modulation levels, transmission powers, OSNR margins}
\Hline{\hspace{-.6 cm}\hrulefill}
\Statex  \hspace{-4mm}\textbf{Stage 1:}RTO (use one of the SPR, SCPR or SCPRR formulations)
\State partition traffic matrix in module of transponder capacity $\mathcal{C}$;
\State route new traffic requests;
\State order routed traffic requests;
\Statex \hspace{-4mm}\textbf{Stage 2:} PSA (use one of the GPSA1, GPSA2, GPSA3, GPSA4, GPSA5, GPSA6 formulations)
\Do
\State solve continuous relaxed GP to obtain $\mathbf{m}, \mathbf{c}$, $\mathbf{p}$, $\bm{\omega}$;
\State $\mathcal{I}\leftarrow 0$;
\Do
\For{all connection requests $q$}
\For{all spectral efficiency values $c$ of Tab. \ref{tab:snr_spc}}
\If {if $c_q$ falls in $\mathcal{I}$-neighborhood of $c$}
\State round and fix $c_q$ to $c$;
\State \textbf{Break};
\EndIf
\EndFor
\EndFor
\State $\mathcal{I} \leftarrow \mathcal{I}+\mathcal{J}$;
\doWhile{there is no valid spectral efficiency value in $\mathbf{c}$}
\State remove rounded values of $\mathbf{c}$;
\doWhile{there is an inacceptable spectral efficiency value in $\mathbf{c}$}
\end{algorithmic}
\end{algorithm}
\section{RTO Sub-Problem}\label{Sec_IV}
SPR is the most familiar routing algorithm that can be formulated as a Binary Linear Program (BLP):
\begin{align}
&\min_{\mathbf{f}}  \sum\limits_{\substack{q \in \mathbf{Q},l \in \mathbf{L}}}f_{q,l} \mathcal{L}_l \label{eq:routing1_g} \\
&\text{s.t.} \sum\limits_{\substack{l \in \mathbf{L},B_l = S_q}}f_{q,l} = 1, \quad \forall q \in \mathbf{Q} \label{eq:routing1_c1}\\
&\text{\quad} \sum\limits_{\substack{l \in \mathbf{L},E_l = D_q}}f_{q,l} = 1, \quad \forall q \in \mathbf{Q} \label{eq:routing1_c2}\\
&\text{\quad} \sum\limits_{\substack{l \in \mathbf{L},E_l = v\\v \neq S_q, v \neq D_q}}f_{q,l}=\sum\limits_{\substack{l \in \mathbf{L},B_l = v\\v \neq S_q, v \neq D_q}}f_{q,l}, \quad \forall q \in \mathbf{Q}, \forall v \in \mathbf{V} \label{eq:routing1_c3}
\end{align}
where $f_{q,l}$ is a binary variable that is $1$ if connection request $q$ is routed over link $l$ and $0$ otherwise. $S_q$ and $D_q$ denote $q$-th connection request source and destination nodes while $B_l$ and $E_l$ show the begin and end nodes of link $l$. The cost function is the route length where $\mathcal{L}_l$ shows $l$-th link length. Constraints \eqref{eq:routing1_c1}-\eqref{eq:routing1_c2} state that traffic should be added and dropped at its source and destination nodes. Constraint  \eqref{eq:routing1_c3} guarantees route continuity.  SPR is suitable when the communication noise is related to route length such as ASE noise. When fiber nonlinearities are important, the cost function should be revised to address the nonlinear interferences. The number of common links, spectrum width and transmission optical power are the main parameters that affect the amount of NLIs \cite{johannisson2014modeling, poggiolini2014gn}. To simultaneously minimize NLIs and ASE noises, one idea is to add minimization of the number of common links to the SPR goal function:
\begin{align}
\quad \sum\limits_{\substack{q,i \in \mathbf{Q},l \in \mathbf{L}}}f_{q,l}  f_{i,l}\mathcal{L}_l \label{eq:routing2_g}
\end{align}
We refer to Binary Quadratic Programming (BQP) formulation of \eqref{eq:routing2_g} and \eqref{eq:routing1_c1}-\eqref{eq:routing1_c3} as SCPR. Traffic volume $\mathcal{R}_q$ can be considered as an indicator of the NLI power since higher traffic volumes need more spectrum width and higher modulation levels and consequently higher transmitted optical power due to required minimum OSNR. Therefore, in SCPRR the goal function is set to:
\begin{align}
\quad \sum\limits_{\substack{q,i \in \mathbf{Q},l \in \mathbf{L}}}f_{q,l}  f_{i,l}\mathcal{R}_i\mathcal{L}_l \label{eq:routing3_g}
\end{align}
Following each routing algorithm, the request traffics are descendingly ordered according to their corresponding cost function. As a result, for SPR, SCPR and SCPRR, traffics are respectively ordered according to cost functions \eqref{eq:routing1_g}, \eqref{eq:routing2_g} and \eqref{eq:routing3_g}.
\section{PSA Sub-Problem}\label{Sec_V}
As mentioned, we formulate PSA as a GP to save run time and reduce optimization complexity. To develop the GPSA formulation, a posynomial expression for OSNR and a posynomial curve fitting for $(c, \Theta(c))$ pairs are required.
\subsection{Posynomial Expressions}\label{Sec_V_A}
The Gaussian noise model considers the NLIs between channels caused by the Kerr effect as additive Gaussian noise and combines this incoherently with ASE noise due to optical fiber amplifiers. The model holds for polarization-multiplexed coherent systems in which equal-length fiber span losses are compensated by optical amplifiers and there is no inline compensation for chromatic dispersion \cite{johannisson2014modeling, poggiolini2014gn}. Nonlinear cross-channel interference for $q$-th connection request $X_q$ (nonlinear kerr effect of connection request $q$ from other connection requests) is expressed as:
\begin{align}
X_{q} = \varsigma p_q \sum\limits_{\substack{i \in \mathbf{Q}\\i \neq q}}\frac{p_{i}^{2}}{\Delta_i^2}\mathcal{N}_{q,i}\log\abs{\frac{\abs{\omega_q-\omega_i}+\frac{\Delta_i}{2}}{\abs{\omega_q-\omega_i}-\frac{\Delta_i}{2}}}, \forall q \in \mathbf{Q}
\end{align}
$\varsigma$ is a constant coefficient and equal to:
\begin{equation}
\varsigma = \frac{3\gamma^2}{2\alpha\pi\abs{\beta_2}}
\end{equation}
where $\alpha$, $\beta_2$ and $\gamma$ are optical fiber attenuation, dispersion and nonlinear constants. Nonlinear self-channel interference power for $q$-th connection request $Y_q$ (nonlinear kerr effect of connection request $q$ on itself) can also be expressed as:
\begin{align}
Y_{q} = \varsigma\mathcal{N}_q  \frac{p_q^3}{\Delta_q^2} \sinh^{-1}(\iota\Delta_q^2), \forall q \in \mathbf{Q}
\end{align}
where $\iota$ is:
\begin{equation}
\iota = \frac{\pi^2\abs{\beta_2}}{2\alpha}
\end{equation}
In addition to the NLIs, each optical amplifier adds white Gaussian noise with power $E_q$:
\begin{equation}
E_{q} = \zeta \mathcal{N}_{q}\Delta_q,\quad \forall q \in \mathbf{Q}
\end{equation}
$\zeta$ is ASE noise coefficient and defined as:
\begin{equation}
\zeta = (e^{\alpha L}-1)h\nu n_{sp}
\end{equation}
where $L$ is the fiber length per span, $n_{sp}$ is the spontaneous emission factor, $\nu$ is the light frequency, and $h$ is Planck’s constant. Combining the linear and nonlinear noises, $q$-th connection request OSNR is \cite{johannisson2014modeling, poggiolini2014gn}:
\begin{equation}
\Psi_{q} = \frac{p_q}{E_{q}+X_{q}+Y_{q}}, \forall q \in \mathbf{Q}
\end{equation}

\begin{figure}[t!] 
\center{\includegraphics[scale=0.45]{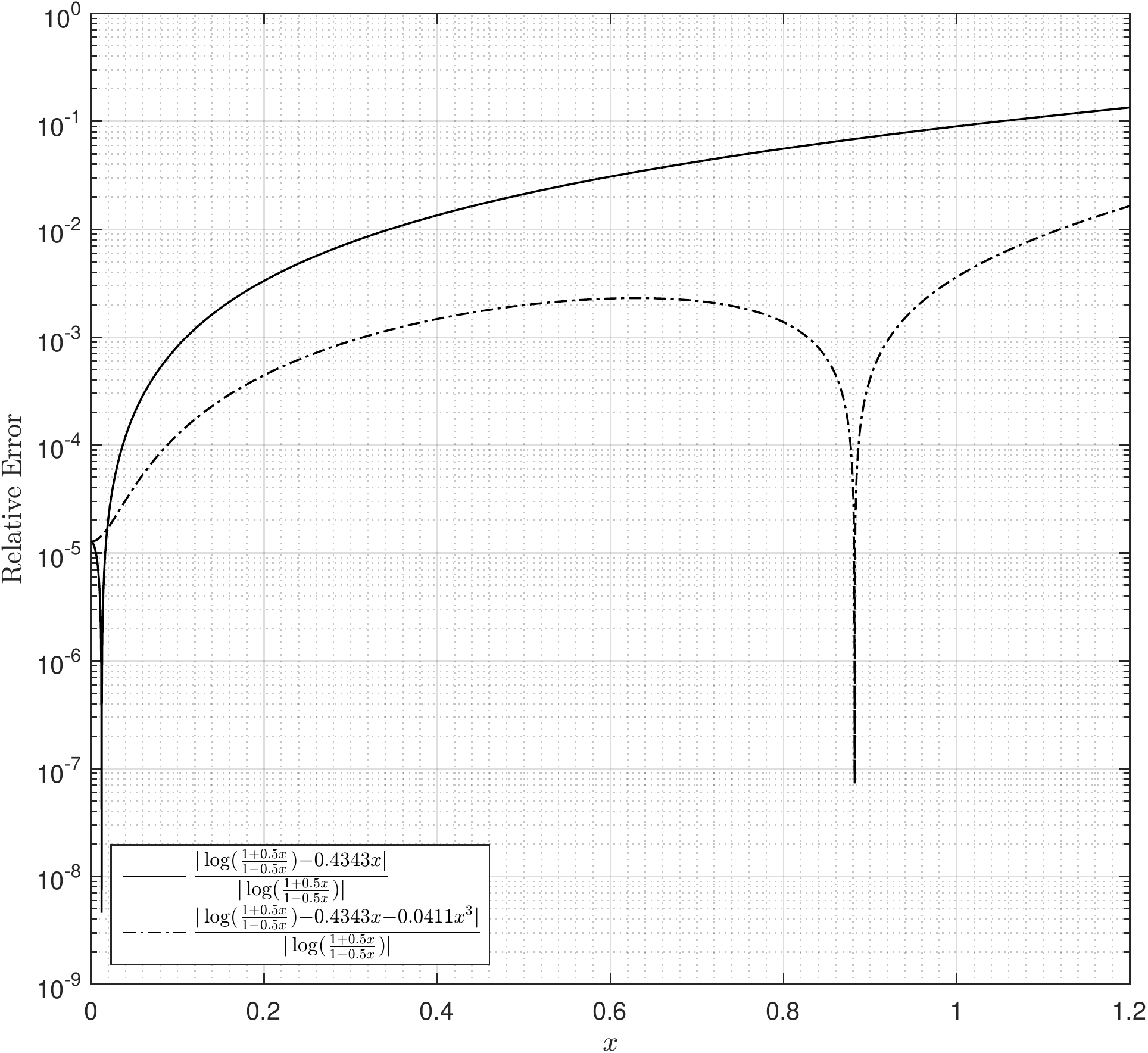}}
\center{\caption{\label{fig:xci} Relative error between function $\log{\Big(\frac{1+0.5x}{1-0.5x}\Big)}$ and its posynomial approximations in \eqref{eq:xci_app1} and \eqref{eq:xci_app2}.}}
\end{figure}

To have a posynomial expression for OSNR, we need to provide posynomial approximations for $\log$ and $\sinh^{-1}$ terms in $X_q$ and $Y_q$, respectively. Now consider the following $\log$ function and its two proposed posynomial approximations:
\begin{align}
&\log{\Big(\frac{1+0.5x}{1-0.5x}\Big)}\approx \kappa_1 x, \quad 0 \leqslant x \leqslant 1.2 \label{eq:xci_app1} \\
&\log{\Big(\frac{1+0.5x}{1-0.5x}\Big)}\approx  \kappa_1 x +  \kappa_2x^3, \quad 0 \leqslant x \leqslant 1.2 \label{eq:xci_app2}
\end{align}
where $ \kappa_1=0.4343$ and $ \kappa_2=0.0411$. Fig. \ref{fig:xci} shows the absolute relative error between the offered values of $\log{\Big(\frac{1+0.5x}{1-0.5x}\Big)}$ and its approximations in \eqref{eq:xci_app1} and \eqref{eq:xci_app2}. Clearly, there is a very good match between the function and its approximations for $0 \leqslant x \leqslant 1.2$. Using the proposed posynomial approximations:
\begin{align}\label{eq:xci_app_expr}
X_{q} & = \varsigma p_q \sum\limits_{\substack{i \in \mathbf{Q}\\i \neq q}}\frac{p_{i}^{2}}{\Delta_i^2}\mathcal{N}_{q,i}\log\abs{\frac{\abs{\omega_q-\omega_i}+\frac{\Delta_i}{2}}{\abs{\omega_q-\omega_i}-\frac{\Delta_i}{2}}}\\
\nonumber &= \varsigma p_q \sum\limits_{\substack{i \in Q\\i \neq q}}\frac{p_{i}^{2}}{\Delta_i^2}\mathcal{N}_{q,i}\log\abs{\frac{1+\frac{\Delta_i}{2d_{q,i}}}{1-\frac{\Delta_i}{2d_{q,i}}}}\\
\nonumber &\approx \varsigma p_q \sum\limits_{\substack{i \in Q\\i \neq q}}\frac{p_{i}^{2}}{\Delta_i^2}\mathcal{N}_{q,i}\Big[\frac{\kappa_1\Delta_i}{d_{q,i}}+\frac{\kappa_2\Delta_i^3}{d_{q,i}^3}\Big]\\
\nonumber &\approx \varsigma p_q \sum\limits_{\substack{i \in Q\\i \neq q}}p_{i}^{2}\mathcal{N}_{q,i}\Big[\frac{\kappa_1}{\Delta_i d_{q,i}}+\frac{\kappa_2\Delta_i}{d_{q,i}^3}\Big]\\
\nonumber &\approx \varsigma p_q \sum\limits_{\substack{i \in Q\\i \neq q}}p_{i}^{2}\mathcal{N}_{q,i}\frac{\kappa_1}{\Delta_id_{q,i}}, \forall q \in \mathbf{Q}
\end{align}
where $d_{q,i}$ is the distance between carrier frequencies $\omega_q$ and $\omega_i$ and equals to $d_{q,i} = \abs{\omega_q-\omega_i}$. Note that for almost all practical scenarios, we have  $0 \leqslant \frac{\Delta_i}{2d_{q,i}} \leqslant 1.2$. For instance, a typical transponder with $\mathcal{R}_i=\mathcal{C}=100$ Gbps \cite{khodakarami2014flexible} needs at most $\Delta_i=50$ GHz bandwidth when it is assigned the lowest modulation format $c_i=2$. Considering a typical value of $\mathcal{G}=20$ GHz \cite{khodakarami2014flexible}, the maximum value of $\frac{\Delta_i}{d_{q,i}}$ is achieved when channels $q$ and $i$ are adjacent and is upper limited by $\frac{50}{20+50/2}=1.11$. Furthermore for non-adjacent channels, the value of  $\frac{\Delta_i}{d_{q,i}}$ is much lower than $1.11$ which is a good news since the accuracy of the approximations is improved for lower values of $\frac{\Delta_i}{2d_{q,i}}$. Unlink the piecewise linear approximated cross-channel interference expression in \cite{yan2017joint} where its relative error is a function of $p_i$, the relative error of \eqref{eq:xci_app_expr} is independent of $p_i$. Furthermore, for fixed value of $p_i$, the relative error of posynomial cross-channel interference is more than one order of magnitude lower than its corresponding value for linear approximation in \cite{yan2017joint}. As a result, the posynomial approximation provides more accurate and reliable results.

For sufficiently small values of $x$, we have $\sinh^{-1}(x) \approx x$. Thus, for practical values of $\beta_2$, $\alpha$ and $\Delta_q$, we can write:
\begin{align}\label{eq:sci_app_expr}
Y_{q} = \varsigma\mathcal{N}_q  \frac{p_q^3}{\Delta_q^2} \sinh^{-1}(\iota\Delta_q^2)\approx\varsigma\iota\mathcal{N}_qp_q^3
, \forall q \in \mathbf{Q}
\end{align}

\begin{figure}[t!] 
\center{\includegraphics[scale=0.45]{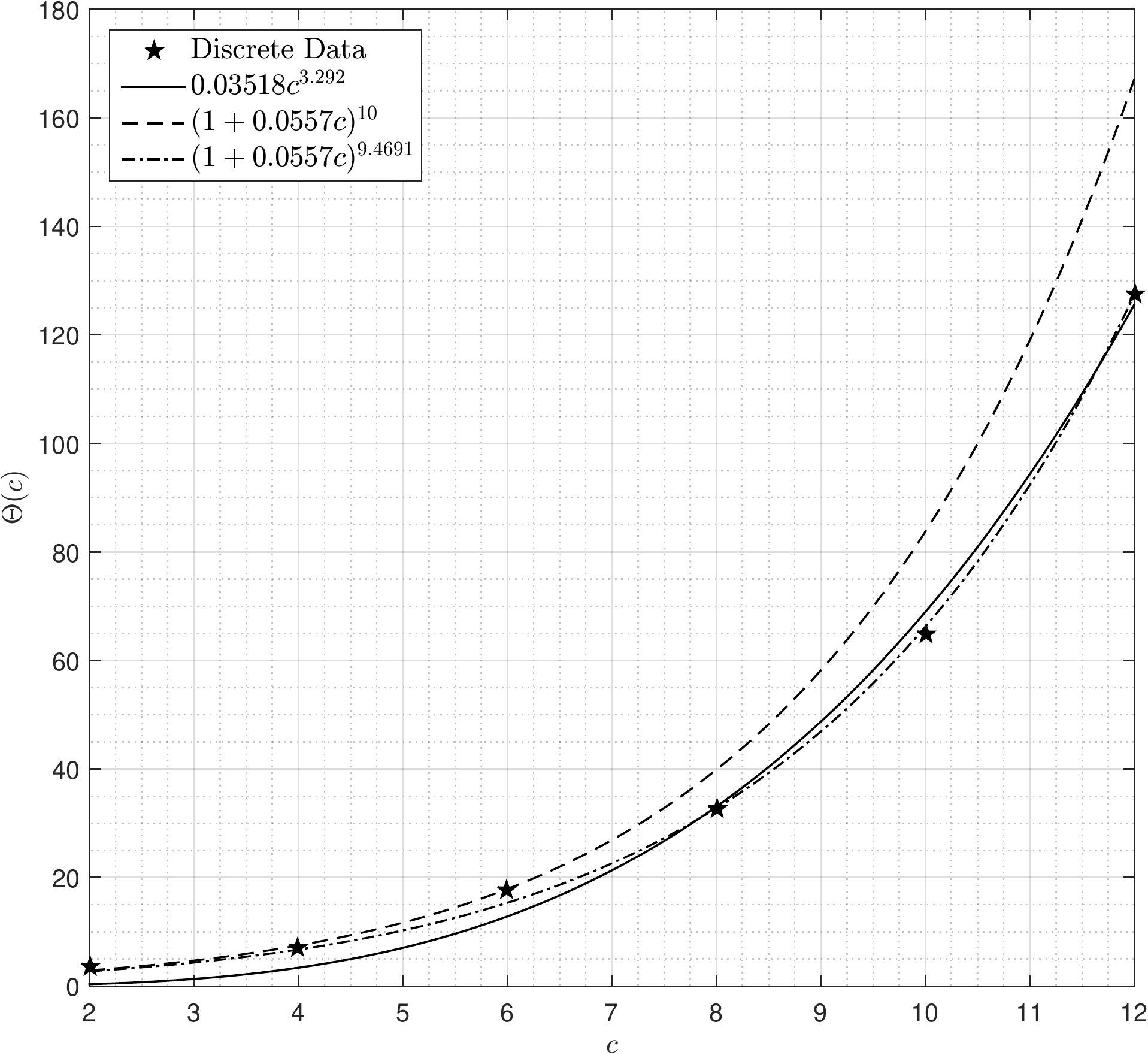}}
\center{\caption{\label{fig:snr} Different curve fittings \eqref{eq:snr_osnr_app1}, \eqref{eq:snr_osnr_app2} and \eqref{eq:snr_osnr_app3} for spectral efficiency-minimum required OSNR samples given in Tab. \ref{tab:snr_spc}.}}
\end{figure}

The available modulation formats along with their corresponding spectral efficiency $c$ and minimum required OSNR $\Theta(c)$ have been collected in Tab. \ref{tab:snr_spc}. We fit the given $(c,\Theta(c))$ samples with the following posynomial expressions:
\begin{align}
& \Theta(c) \approx \kappa_3c^{\kappa_4}, \quad 2 \leqslant c \leqslant 12 \label{eq:snr_osnr_app1}\\
& \Theta(c) \approx (1+\kappa_5 c)^{\kappa_6}, \quad 2 \leqslant c \leqslant 12 \label{eq:snr_osnr_app2}\\
& \Theta(c) \approx (1+\kappa_5c)^{\kappa_7}, \quad 2 \leqslant c \leqslant 12 \label{eq:snr_osnr_app3} 
\end{align}
where $\kappa_3=0.0351$, $\kappa_4=3.292$, $\kappa_5=0.0557$, $\kappa_6=10$ and $\kappa_7=9.4691$. Fig. \ref{fig:snr} show data samples and three posynomial expressions for data curve fitting. The curve fitting \eqref{eq:snr_osnr_app3} has the lowest relative error and the error increases for \eqref{eq:snr_osnr_app2} and \eqref{eq:snr_osnr_app1}, respectively.

\subsection{Problem Formulation}\label{Sec_V_B}
Here we provide a concise review of the MINLP formulation for PSA and then introduce our GPSA counterparts. A MINLP formulation for PSA is as follows \cite{yan2015resource}:
\begin{align}
&\min_{\mathbf{c}, \mathbf{m}, \bm{\omega}, \mathbf{p}, \tau} \quad \mathcal{K}_1\tau+ \mathcal{K}_2\sum\limits_{q\in \mathbf{Q}}p_q+ \mathcal{K}_3 \sum\limits_{q\in \mathbf{Q}}m_q^{-1}\label{eq:nonlinear_g}\\
&\text{s.t.} \quad \Psi_q \geqslant m_q \Theta_q, \quad \forall q \in \mathbf{Q} \label{eq:nonlinear_c1}\\
\nonumber & \text{\quad}\quad \omega_{\Upsilon_{l, j}}+\frac{\Delta_{\Upsilon_{l, j}}}{2} + \mathcal{G}\leqslant \omega_{\Upsilon_{l, j+1}}-\frac{\Delta_{\Upsilon_{l, j+1}}}{2}, \quad \forall l \in \mathbf{L} \\
&\text{\quad} \quad, j = 1, \cdots, \abs{\mathbf{Q}_l}-1 \label{eq:nonlinear_c2}\\
&\text{\quad} \quad \frac{\Delta_q}{2} + \omega_q \leqslant \tau, \quad \forall q \in \mathbf{Q} \label{eq:nonlinear_c3}\\
&\text{\quad} \quad m_q \geqslant \mathcal{M}, \quad \forall q \in \mathbf{Q} \label{eq:nonlinear_c4}\\
&\text{\quad} \quad \tau \leqslant \mathcal{B} \label{eq:nonlinear_c5}
\end{align}
where $\tau$ is the penalty term for occupied frequency upper bound and $\mathcal{M}$ is minimum OSNR margin. $\Upsilon_{l, j}$ is a function that shows which connection request occupies $j$-th channel on link $l$ and its values are determined during RTO sub-problem. Constraint \eqref{eq:nonlinear_c1} is the QoS constraint that forces $q$-th connection request OSNR to be at least  $m_q$ times greater than its required minimum OSNR where $m_q$ is an optimization variable that shows OSNR margin of $q$-th connection request. Constraint \eqref{eq:nonlinear_c2} is nonoverlapping-guard constraint that prevents two connections requests to share a same frequency spectrum. It also guarantees the required guard band between any two adjacent connection requests. Constraint \eqref{eq:nonlinear_c3} compresses the used spectrum to decrease the penalty term for occupied frequency upper bound. Constraint \eqref{eq:nonlinear_c4} holds all OSNR margins greater than its minimum value $\mathcal{M}$ and finally, constraint \eqref{eq:nonlinear_c5} blocks spectrum assignment outside the fiber bandwidth $\mathcal{B}$. The formulation is flexible enough to minimize the maximum bandwidth usage on all the links (to increase remaining resources for more traffic demands), the total transmitted optical power (to reduce nonlinearity in the fibers) and maximize the total SNR margins (to improve the robustness and capacity of the network) \cite{yan2015resource}. 

Now, consider the simplest posynomial approximations, \ie \eqref{eq:xci_app1} and \eqref{eq:snr_osnr_app1} and note the following GP:
\begin{align}
&\min_{\mathbf{d}, \mathbf{c}, \bm{\omega}, \mathbf{p}, \tau} \mathcal{K}_1\tau+\mathcal{K}_2\sum\limits_{q\in \mathbf{Q}} p_q+\mathcal{K}_3 \sum\limits_{q\in \mathbf{Q}}m_q^{-1}+\mathcal{K}_4\sum\limits_{\substack{q,i\in \mathbf{Q}\\i \neq q\\ \mathcal{N}_{q,i} \neq 0}}d_{q,i}^{-1} \label{eq:gp_1_g} \\
\nonumber &\text{s.t.} \quad \kappa_3\zeta m_q\mathcal{N}_{q}\mathcal{R}_{q}c_{q}^{\kappa_4-1}p_{q}^{-1}+\kappa_3\varsigma\iota m_q \mathcal{N}_{q}p_{q}^2c_q^{\kappa_4}+\\
& \text{\quad}\quad  \kappa_3\kappa_1 \varsigma m_q c_q^{\kappa_4}\sum\limits_{\substack{i \in \mathbf{Q}\\i \neq q}}p_{i}^{2}\mathcal{N}_{q,i}\mathcal{R}_i^{-1} c_i d_{q,i}^{-1}\leqslant 1, \quad \forall q \in \mathbf{Q} \label{eq:gp_1_c1} \\
\nonumber & \text{\quad} \quad \omega_{\Upsilon_{l, j}}\omega_{\Upsilon_{l, j+1}}^{-1}+0.5\mathcal{R}_{\Upsilon_{l, j}} c_{\Upsilon_{l, j}}^{-1} \omega_{\Upsilon_{l, j+1}}^{-1}+\mathcal{G}\omega_{\Upsilon_{l, j+1}}^{-1}\\
\nonumber & \text{\quad}\quad +0.5\mathcal{R}_{\Upsilon_{l, j+1}} c_{\Upsilon_{l, j+1}}^{-1} \omega_{\Upsilon_{l, j+1}}^{-1}\leqslant 1\\
& \text{\quad}\quad , \quad \forall l \in \mathbf{L}, j = 1, \cdots, \abs{\mathbf{Q}_l}-1 \label{eq:gp_1_c2}  \\
&\text{\quad} \quad 0.5\mathcal{R}_q c_q^{-1}\tau^{-1} + \omega_q\tau^{-1} \leqslant 1, \quad \forall q \in \mathbf{Q} \label{eq:gp_1_c3} \\
& \text{\quad}\quad \mathcal{M} m_q^{-1} \leqslant 1, \quad \forall q \in \mathbf{Q} \label{eq:gp_1_c4}  \\
& \text{\quad}\quad \tau \mathcal{B}^{-1} \leqslant 1 \label{eq:gp_1_c5}  \\
\nonumber &\text{\quad} \quad d_{\Upsilon_{l, i},\Upsilon_{l, j}} \omega_{\Upsilon_{l, i}}^{-1}+ \omega_{\Upsilon_{l, j}}\omega_{\Upsilon_{l, i}}^{-1}\leqslant  1,\quad \forall l \in \mathbf{L} \label{eq:gp_1_c6} \\
&\text{\quad} \quad, j = 1, \cdots, \abs{\mathbf{Q}_l}-1, i = j+1, \cdots, \abs{\mathbf{Q}_l}\\
\nonumber &\text{\quad} \quad d_{\Upsilon_{l, i},\Upsilon_{l, j}}\omega_{\Upsilon_{l, j}}^{-1}+\omega_{\Upsilon_{l, i}}\omega_{\Upsilon_{l, j}}^{-1} \leqslant  1 ,\quad \forall l \in \mathbf{L} \\
&\text{\quad} \quad, j = 2, \cdots, \abs{\mathbf{Q}_l}, i = 1, \cdots, j-1 \label{eq:gp_1_c7} 
\end{align}
Ignoring constraints of \eqref{eq:gp_1_c6} and \eqref{eq:gp_1_c7} and the fourth penalty term of the goal function \eqref{eq:gp_1_g}, the above formulation is equivalent GP of the previous MINLP in which posynomial expressions of \eqref{eq:snr_osnr_app1} and \eqref{eq:xci_app1} have been used for QoS constraint \eqref{eq:gp_1_c1}. The added constraints and penalty term are to guarantee the equality of $d_{q,i} = \abs{\omega_q-\omega_i}$. In fact, constraints \eqref{eq:gp_1_c6} and \eqref{eq:gp_1_c7} specifies that $d_{q,i} \leqslant \abs{\omega_q-\omega_i}$ but added posynomial penalty term $k_4\sum d_{q,i}^{-1}$ forces $d_{q,i}$ takes their maximum values, \ie $\abs{\omega_q-\omega_i}$. Also this GP formulation has more variables and constraints compared to its MINLP counterpart but its solution is rapidly achieved using convexification techniques and fast convex optimization algorithms. 

To improve the accuracy of the proposed formulation, one can use more accurate posynomial approximations. For \eqref{eq:snr_osnr_app1} and \eqref{eq:xci_app2}, the previous formulation is valid if we replace \eqref{eq:gp_1_c1} with the following:
\begin{align}
\nonumber & \kappa_3\zeta m_q \mathcal{N}_{q}\mathcal{R}_{q}c_{q}^{\kappa_4-1}p_{q}^{-1}+\kappa_3\varsigma\iota m_q \mathcal{N}_{q}p_{q}^2c_q^{\kappa_4}\\
 &  + \kappa_3\kappa_1 \varsigma m_q c_q^{\kappa_4}\sum\limits_{\substack{i \in \mathbf{Q}\\i \neq q}}p_{i}^{2}\mathcal{N}_{q,i}\mathcal{R}_i^{-1} c_i d_{q,i}^{-1} \label{eq:gp_2_c1}\\
\nonumber &  + \kappa_3\kappa_2 \varsigma m_q c_q^{\kappa_4}\sum\limits_{\substack{i \in \mathbf{Q}\\i \neq q}}p_{i}^{2}\mathcal{N}_{qi}\mathcal{R}_ic_i ^{-1} d_{q,i}^{-3}\leqslant 1 , \quad \forall q \in \mathbf{Q}  
\end{align}
For posynomial expressions \eqref{eq:snr_osnr_app2} and \eqref{eq:xci_app1}, the constraint \eqref{eq:gp_1_c1}  should be replaced with:
\begin{align}
\nonumber & m_q (1+\kappa_5c_q)^{\kappa_6}\Big[\zeta\mathcal{N}_{q}\mathcal{R}_{q}c_{q}^{-1}p_{q}^{-1}+\varsigma\iota \mathcal{N}_{q}p_{q}^2\\
 &+ \kappa_1\varsigma\sum\limits_{\substack{i \in \mathbf{Q}\\i \neq q}}p_{i}^{2}\mathcal{N}_{q,i}\mathcal{R}_i^{-1} c_i d_{q,i}^{-1}\Big]\leqslant 1, \quad \forall q \in \mathbf{Q} \label{eq:gp_3_c1} 
\end{align}
while for approximations \eqref{eq:snr_osnr_app2} and \eqref{eq:xci_app2}:
\begin{align}
\nonumber &m_q(1+\kappa_5c_q)^{\kappa_6}\Big[\zeta\mathcal{N}_{q}\mathcal{R}_{q}c_{q}^{-1}p_{q}^{-1}+\varsigma\iota \mathcal{N}_{q}p_{q}^2\\
\nonumber &+ \kappa_1 \varsigma\sum\limits_{\substack{i \in \mathbf{Q}\\i \neq q}}p_{i}^{2}\mathcal{N}_{q,i}\mathcal{R}_i^{-1} c_i d_{q,i}^{-1}\\
&  + \kappa_2 \varsigma \sum\limits_{\substack{i \in \mathbf{Q}\\i \neq q}}p_{i}^{2}\mathcal{N}_{q,i}\mathcal{R}_ic_i ^{-1} d_{q,i}^{-3}\Big]\leqslant 1, \quad \forall q \in \mathbf{Q} \label{eq:gp_4_c1} 
\end{align}
Note that in \eqref{eq:gp_3_c1} and \eqref{eq:gp_4_c1}, Newton binomial expansion of the term $(1+\kappa_5c_q)^{\kappa_6}$ should be multiplied by the remaining posynomial terms and the result is clearly posynomial. This fact is not true for \eqref{eq:snr_osnr_app3} in which $\kappa_7$ is not a natural number. Therefore, auxiliary variable $t_q$'s should be used to change the non-posynomial multiplier of $(1+\kappa_5c_q)^{\kappa_7}$ to a posynomial expression \cite{boyd2007tutorial}. Consequently for \eqref{eq:snr_osnr_app3} and \eqref{eq:xci_app1}, we should change constraint \eqref{eq:gp_1_c1} to:
\begin{align}
\nonumber &m_q t_q^{\kappa_7}\Big[\zeta\mathcal{N}_{q}\mathcal{R}_{q}c_{q}^{-1}p_{q}^{-1}+\varsigma\iota \mathcal{N}_{q}p_{q}^2\\
 &+ \kappa_1 \varsigma\sum\limits_{\substack{i \in \mathbf{Q}\\i \neq q}}p_{i}^{2}\mathcal{N}_{q,i}\mathcal{R}_i^{-1} c_i d_{q,i}^{-1}\Big]\leqslant 1, \quad \forall q \in \mathbf{Q} \label{eq:gp_5_c1} 
\end{align}
and add a new constraint:
\begin{equation} \label{eq:gp_5_c8} 
 t_q^{-1}+\kappa_5c_qt_q^{-1} \leqslant 1, \quad \forall q \in \mathbf{Q} 
\end{equation}
Finally for approximations \eqref{eq:snr_osnr_app3} and \eqref{eq:xci_app2}, we similarly add constraint \eqref{eq:gp_5_c8} and change the QoS constraint to:
\begin{align}
\nonumber &m_q t_q^{\kappa_7}\Big[\zeta\mathcal{N}_{q}\mathcal{R}_{q}c_{q}^{-1}p_{q}^{-1}+\varsigma\iota \mathcal{N}_{q}p_{q}^2+ \kappa_1 \varsigma\sum\limits_{\substack{i \in \mathbf{Q}\\i \neq q}}p_{i}^{2}\mathcal{N}_{q,i}\mathcal{R}_i^{-1} c_i d_{q,i}^{-1}\\
&  + \kappa_2 \varsigma \sum\limits_{\substack{i \in \mathbf{Q}\\i \neq q}}p_{i}^{2}\mathcal{N}_{q,i}\mathcal{R}_ic_i ^{-1} d_{q,i}^{-3}\Big]\leqslant 1, \quad \forall q \in \mathbf{Q} \label{eq:gp_6_c1} 
\end{align}

\section{Numerical Results}\label{Sec_VI}
\begin{figure}[t!] 
\center{\includegraphics[scale=0.45]{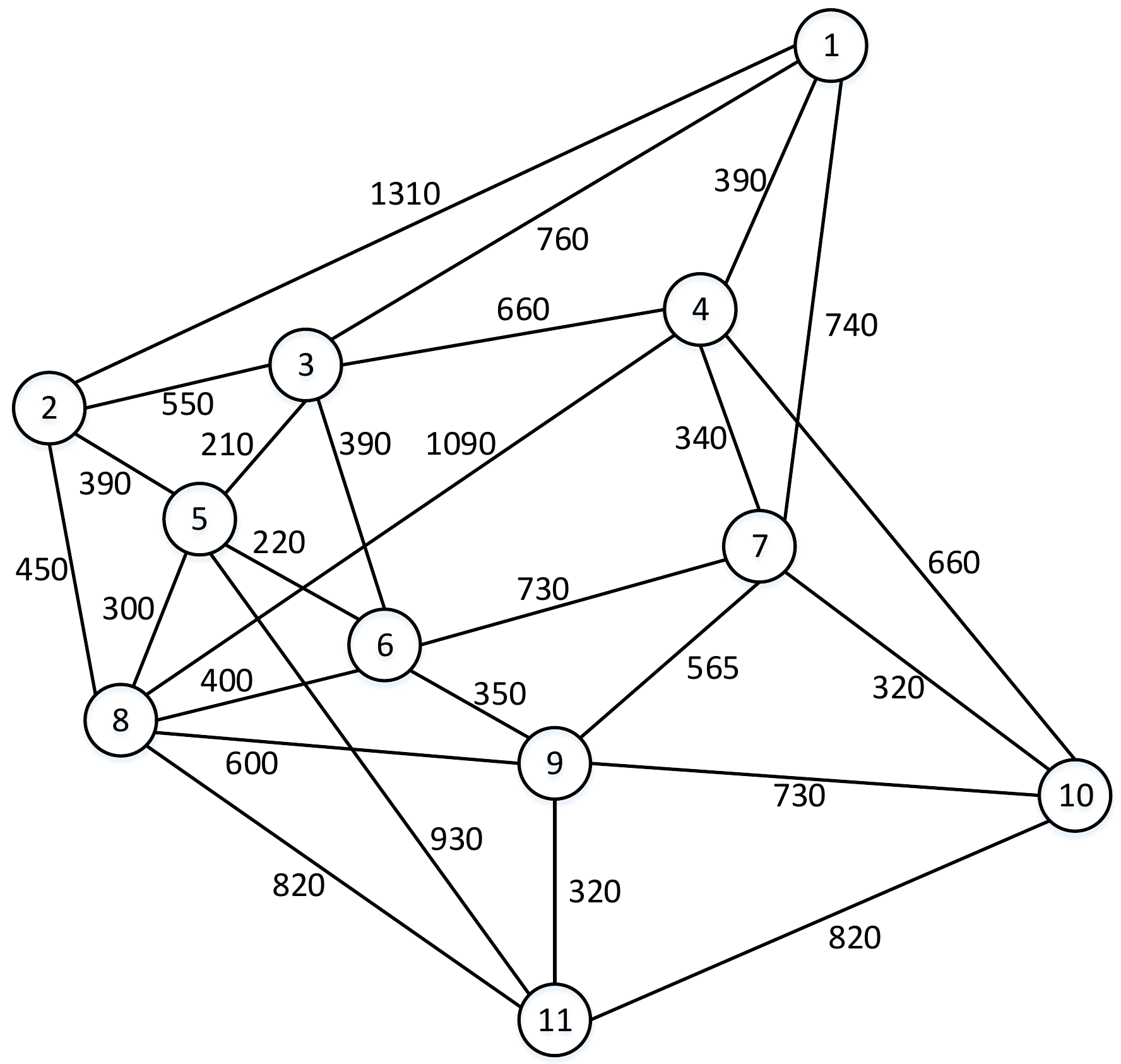}}
\center{\caption{\label{fig:euronet} European Cost239 optical network with $11$ optical nodes and $26$ bi-directional links. Digits on each link shows its length in km.}}
\end{figure}

In this section, we use simulation results to demonstrate the performance of the proposed formulations and algorithms. The European Cost239 optical network is considered with the topology and the normalized traffic matrix given by Fig. \ref{fig:euronet} and Tab. \ref{tab:traffic_matrix}, respectively \cite{muhammad2014introducing, khodakarami2014flexible}. The optical link parameters and other constant values are reported in Tab. \ref{tab:sim_param}. We consider the MINLP formulation of \eqref{eq:nonlinear_g}-\eqref{eq:nonlinear_c5} as a benchmark for evaluating the heuristic and formulations \cite{yan2015resource}. Various formulations for RTO and PSA are summarized in Tab. \ref{tab:forms} and their corresponding number of variables and constraints are specified. 
\begin{table}[t!]
\centering
\caption{European Cost239 optical network normalized traffic matrix.}\label{tab:traffic_matrix}
\begin{tabular}{cccccccccccc}
\hline
\hline
\textbf{Node} & $\mathbf{01}$ & $\mathbf{02}$ & $\mathbf{03}$ & $\mathbf{04}$ & $\mathbf{05}$ & $\mathbf{06}$ & $\mathbf{07}$ & $\mathbf{08}$ & $\mathbf{09}$ & $\mathbf{10}$ & $\mathbf{11}$\\
\hline
$\mathbf{01}$ & $00$ & $01$ & $01$ & $03$ & $01$ & $01$ & $01$ & $35$ & $01$ & $01$ & $01$\\
$\mathbf{02}$ & $01$ & $00$ & $05$ & $14$ & $40$ & $01$ & $01$ & $10$ & $03$ & $02$ & $03$\\
$\mathbf{03}$ & $01$ & $05$ & $00$ & $16$ & $24$ & $01$ & $01$ & $05$ & $03$ & $01$ & $02$\\
$\mathbf{04}$ & $03$ & $14$ & $16$ & $00$ & $06$ & $02$ & $02$ & $21$ & $81$ & $09$ & $09$\\
$\mathbf{05}$ & $01$ & $40$ & $24$ & $06$ & $00$ & $01$ & $11$ & $06$ & $11$ & $01$ & $02$\\
$\mathbf{06}$ & $01$ & $01$ & $01$ & $02$ & $01$ & $00$ & $01$ & $01$ & $01$ & $01$ & $01$\\
$\mathbf{07}$ & $01$ & $01$ & $01$ & $02$ & $11$ & $01$ & $00$ & $01$ & $01$ & $01$ & $01$\\
$\mathbf{08}$ & $35$ & $10$ & $05$ & $21$ & $06$ & $01$ & $01$ & $00$ & $06$ & $02$ & $05$\\
$\mathbf{09}$ & $01$ & $03$ & $03$ & $81$ & $11$ & $01$ & $01$ & $06$ & $00$ & $51$ & $06$\\
$\mathbf{10}$ & $01$ & $02$ & $01$ & $09$ & $01$ & $01$ & $01$ & $02$ & $51$ & $00$ & $81$\\
$\mathbf{11}$ & $01$ & $03$ & $02$ & $09$ & $02$ & $01$ & $01$ & $05$ & $06$ & $81$ & $00$\\
\hline
\hline
\end{tabular}
\end{table}
\begin{table*}[t!]
\centering
\caption{Simulation constant parameters.}\label{tab:sim_param}
\begin{tabular}{cccccccccccc}
\hline
\hline
$\abs{\substack{\beta_2}}$ & $\alpha$ & $L$ & $\nu$ & $n_{sp}$ & $\gamma$ & $\mathcal{G}$ & $\mathcal{B}$ & $\mathcal{C}$  & $\mathcal{J}$\\
$\text{fs}^2/\text{m}$ & dB/km & km & THz &  & 1/W/km & GHz & THz & Gbps & \\
\hline
$20393$ & $0.22$ & $80$ & $193.55$ & $1.58$ & $1.3$ & $20$ & $2$ & $100$ & 0.1\\
\hline
\hline
\end{tabular}
\end{table*}
\begin{table*}[t!]
\centering
\caption{ Various formulations for RTO and PSA considered in the paper and their corresponding number of variables and constraints.}\label{tab:forms}
\begin{tabular}{cccccc}
\hline
\hline
Formulation & Type & Goal & Constraint & \#Variables & \#Constraints\\
\hline
SPR & BLP  & \eqref{eq:routing1_g}  & \eqref{eq:routing1_c1}-\eqref{eq:routing1_c3} & $\abs{\mathbf{Q}}\abs{\mathbf{L}}$ & $2\abs{\mathbf{Q}}+\abs{\mathbf{Q}}\abs{\mathbf{V}}$ \\
SCPR & BQP  & \eqref{eq:routing2_g} & \eqref{eq:routing1_c1}-\eqref{eq:routing1_c3} & $\abs{\mathbf{Q}}\abs{\mathbf{L}}$ & $2\abs{\mathbf{Q}}+\abs{\mathbf{Q}}\abs{\mathbf{V}}$\\
SCPRR & BQP  & \eqref{eq:routing3_g}  & \eqref{eq:routing1_c1}-\eqref{eq:routing1_c3} & $\abs{\mathbf{Q}}\abs{\mathbf{L}}$ & $2\abs{\mathbf{Q}}+\abs{\mathbf{Q}}\abs{\mathbf{V}}$ \\
MINLPPSA & MINLP  & \eqref{eq:nonlinear_g} & \eqref{eq:nonlinear_c1}-\eqref{eq:nonlinear_c5} & $4\abs{\mathbf{Q}}+1$ & $3\abs{\mathbf{Q}}+\abs{\mathbf{Q}}\abs{\mathbf{L}}+1$ \\
GPSA1 & MIGP  & \eqref{eq:gp_1_g}  & \eqref{eq:gp_1_c1}-\eqref{eq:gp_1_c7} & $\abs{\mathbf{Q}}^2+4\abs{\mathbf{Q}}+1$ & $3\abs{\mathbf{Q}}+3\abs{\mathbf{Q}}\abs{\mathbf{L}}+1$ \\
GPSA2 & MIGP & \eqref{eq:gp_1_g}  &  \eqref{eq:gp_2_c1}, \eqref{eq:gp_1_c2}-\eqref{eq:gp_1_c7} & $\abs{\mathbf{Q}}^2+4\abs{\mathbf{Q}}+1$ & $3\abs{\mathbf{Q}}+3\abs{\mathbf{Q}}\abs{\mathbf{L}}+1$  \\
GPSA3 & MIGP  & \eqref{eq:gp_1_g}  &  \eqref{eq:gp_3_c1}, \eqref{eq:gp_1_c2}-\eqref{eq:gp_1_c7} & $\abs{\mathbf{Q}}^2+4\abs{\mathbf{Q}}+1$ & $3\abs{\mathbf{Q}}+3\abs{\mathbf{Q}}\abs{\mathbf{L}}+1$ \\
GPSA4 & MIGP  & \eqref{eq:gp_1_g}  &  \eqref{eq:gp_4_c1}, \eqref{eq:gp_1_c2}-\eqref{eq:gp_1_c7} & $\abs{\mathbf{Q}}^2+4\abs{\mathbf{Q}}+1$ & $3\abs{\mathbf{Q}}+3\abs{\mathbf{Q}}\abs{\mathbf{L}}+1$\\
GPSA5 & MIGP & \eqref{eq:gp_1_g} & \eqref{eq:gp_5_c1}, \eqref{eq:gp_1_c2}-\eqref{eq:gp_1_c7}, \eqref{eq:gp_5_c8} & $\abs{\mathbf{Q}}^2+5\abs{\mathbf{Q}}+1$ & $4\abs{\mathbf{Q}}+3\abs{\mathbf{Q}}\abs{\mathbf{L}}+1$\\
GPSA6 & MIGP  & \eqref{eq:gp_1_g} & \eqref{eq:gp_6_c1}, \eqref{eq:gp_1_c2}-\eqref{eq:gp_1_c7}, \eqref{eq:gp_5_c8} & $\abs{\mathbf{Q}}^2+5\abs{\mathbf{Q}}+1$ & $4\abs{\mathbf{Q}}+3\abs{\mathbf{Q}}\abs{\mathbf{L}}+1$ \\
\hline
\hline
\end{tabular}
\end{table*}

\begin{figure}[t!] 
\center{\includegraphics[scale=0.45]{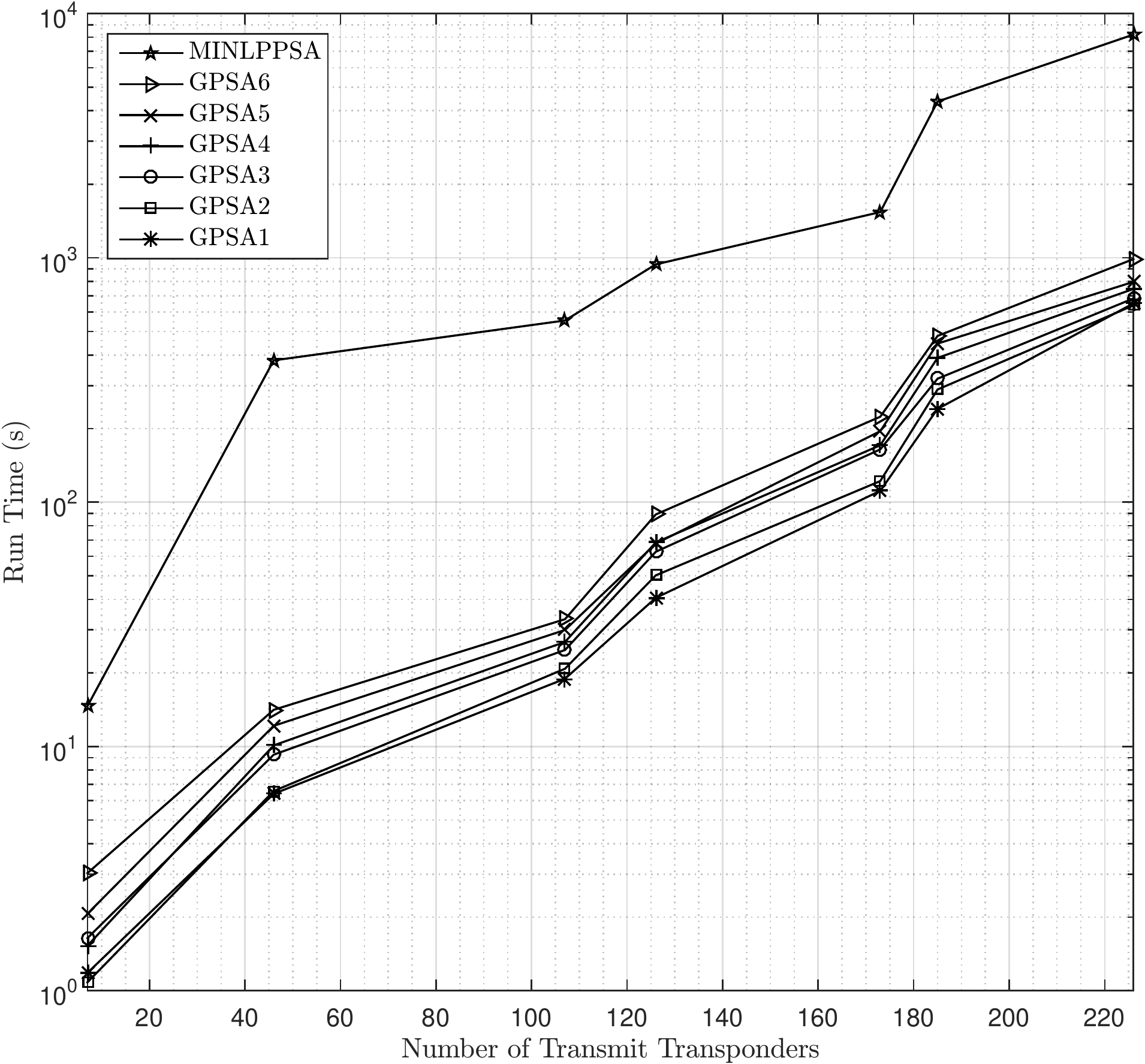}}
\center{\caption{\label{fig:time}Run time in terms of the number of transmit transponders for MINLP and GP formulations of PSA stage of Alg. \ref{alg:rsa}.}}
\end{figure}

Fig. \ref{fig:time} compares the run time of the different PSA formulations for various number of transmit transponders. For all of the formulations, the same RTO (here SPR) are applied. Traffics are randomly selected from traffic matrix in Tab. \ref{tab:traffic_matrix}. GPSAs are modeled and solved using YALMIP \cite{lofberg2005yalmip} and MOSEK \cite{mosek2010mosek} software packages while we use BONMIN \cite{bonami2007bonmin} for solving MINLP PSA (MINLPPSA). Simulations run over a Corei7 8 GB RAM computer. Clearly, the run time is considerably improved for GPSA formulations. Among GPSAs, the run time increases for higher precision posynomial expressions and the best time is obtained for GPSA1, the simplest GPSA. Based on Fig. \ref{fig:time}, the run time of GPSA1 is more than $59$ times shorter than MINLPPSA when there are $46$ transmit transponders. Although GPSA formulations have more number of constraints and variables, their special properties result in very short convergence time.

\begin{figure}[t!] 
\center{\includegraphics[scale=0.45]{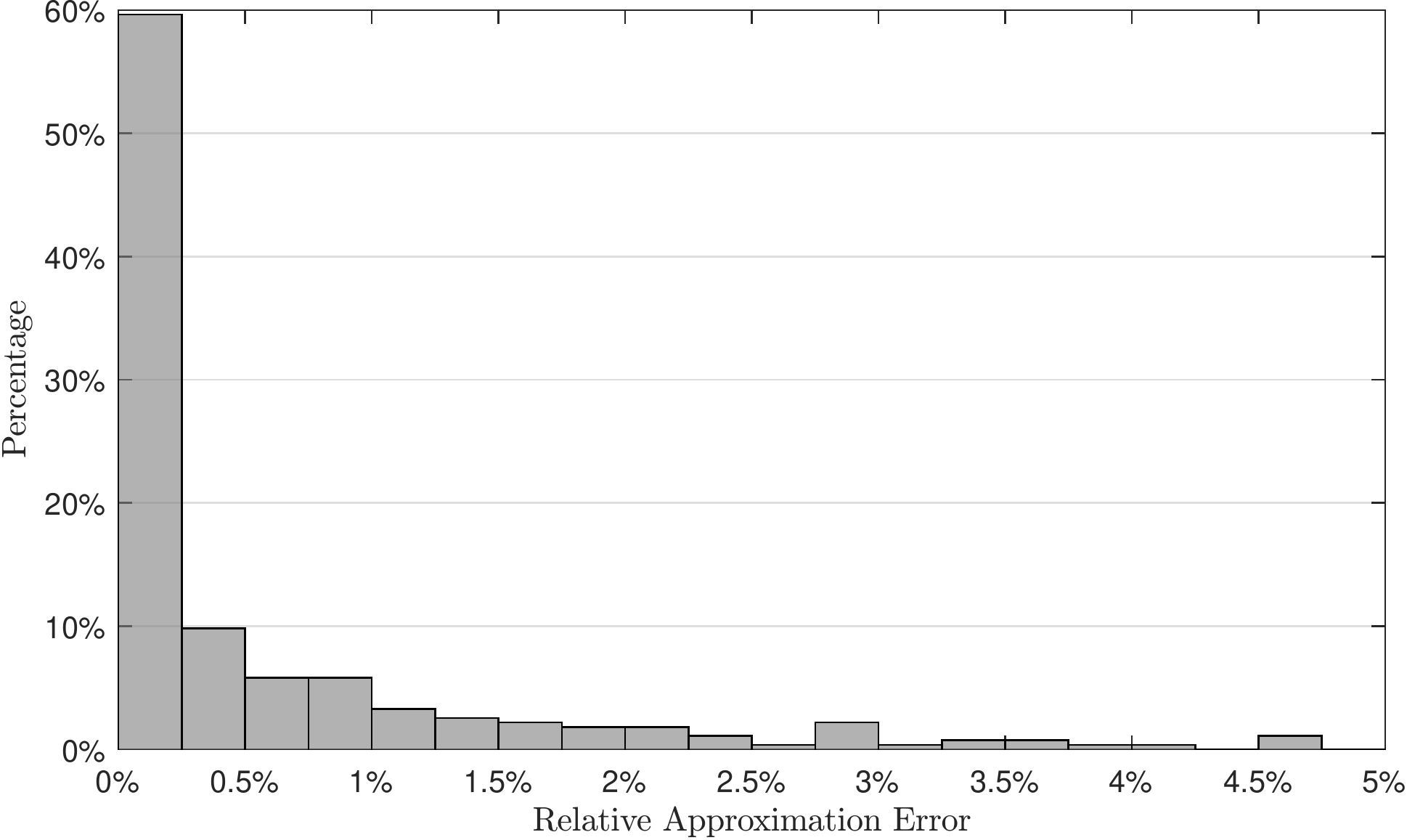}}
\center{\caption{\label{fig:err1} Distribution of OSNR relative errors of GPSA5 with respect to MINLPPSA in European Cost239 optical network with $46$ transmit transponders.}}
\end{figure}

\begin{figure}[t!] 
\center{\includegraphics[scale=0.45]{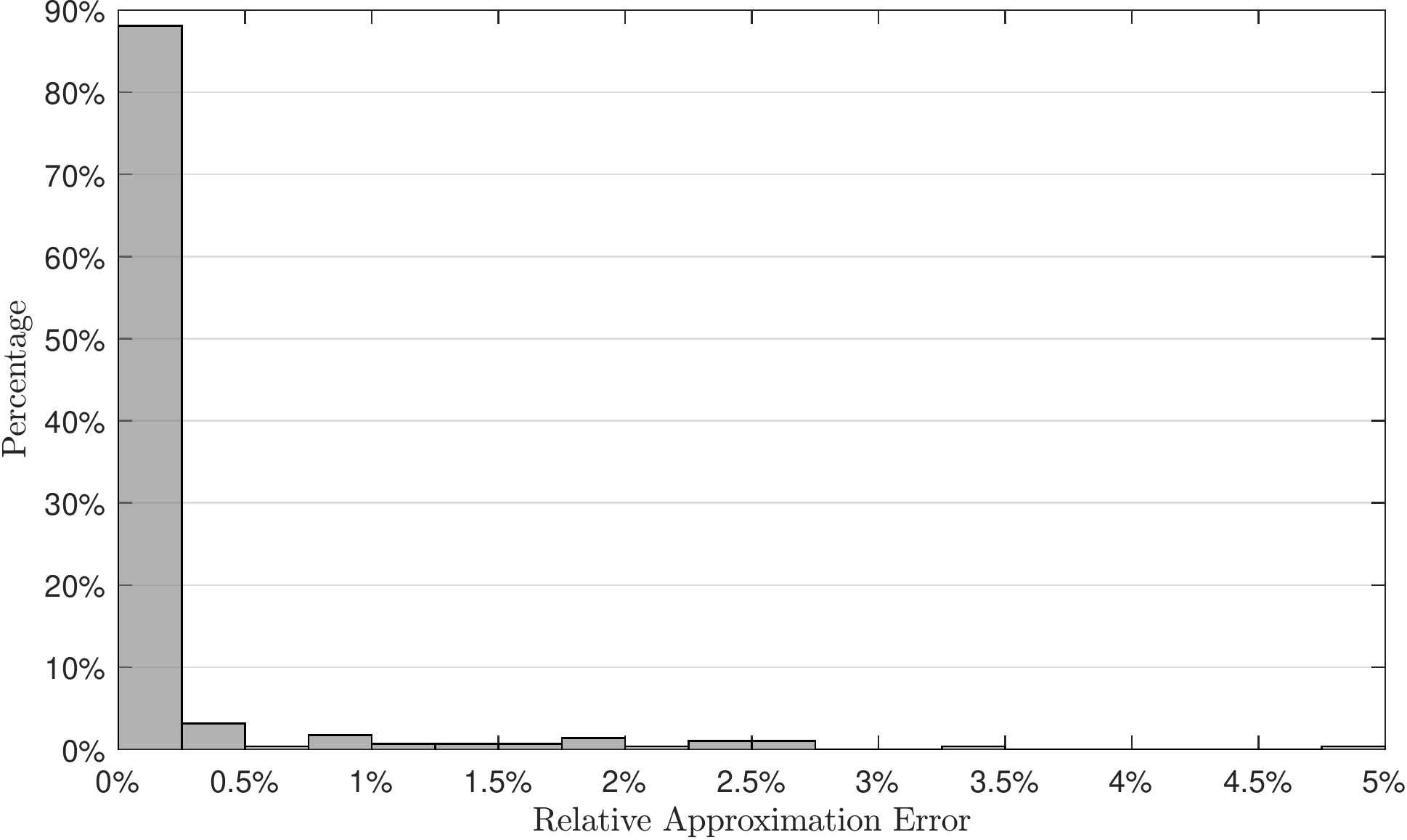}}
\center{\caption{\label{fig:err2} Distribution of OSNR relative errors of GPSA6 with respect to MINLPPSA in European Cost239 optical network with $46$ transmit transponders.}}
\end{figure}

For the European Cost239 optical network in Fig. \ref{fig:euronet} with $46$ transmit transponders, the distribution of OSNR relative errors (with respect to MINLPPSA) given by GPSA5 and GPSA6 solutions are shown in Fig. \ref{fig:err1} and Fig. \ref{fig:err2}, respectively.  The average relative error is $2.13\%$ for GPSA5 and $1.09\%$ for GPSA6. This shows that GPSA6 and GPSA5 provide accurate results with a much shorter run time compared to MINLPPSA.

\begin{figure}[t!] 
\center{\includegraphics[scale=0.45]{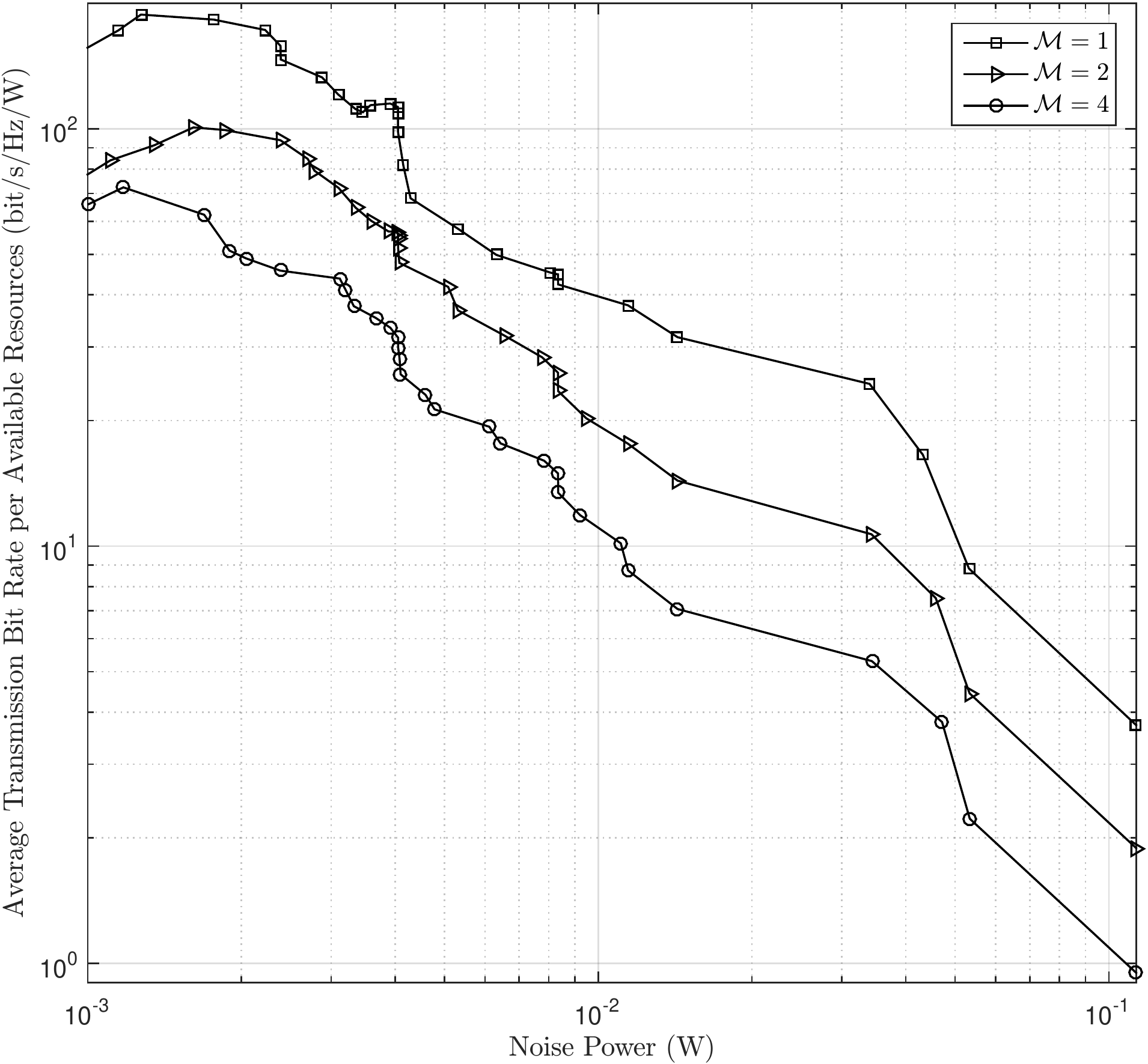}}
\center{\caption{\label{fig:spectral} Average transmission bit rate per available resources $\frac{\mathcal{R}_q}{p_q\Delta_q}$  in terms of total NLIs and ASE noise $X_q+Y_q+E_q$ for different values of OSNR margin $\mathcal{M} = 1, 2, 4$ in European Cost239 optical network with $46$ transmit transponders.}}
\end{figure}

Considering the product of $p_q\Delta_q$ as available communication resources for $q$-th connection request, Fig.  \ref{fig:spectral} shows the average transmission bit rate per available resources $\frac{\mathcal{R}_q}{p_q\Delta_q}$  in terms of total NLIs and ASE noise $X_q+Y_q+E_q$ for different values of minimum OSNR margin $\mathcal{M} = 1, 2, 4$ in European Cost239 optical network with $46$ transmit transponders. As it is seen, when the system noise is increased, the resource utilization of the network is decreased.

\begin{figure}[t!] 
\center{\includegraphics[scale=0.45]{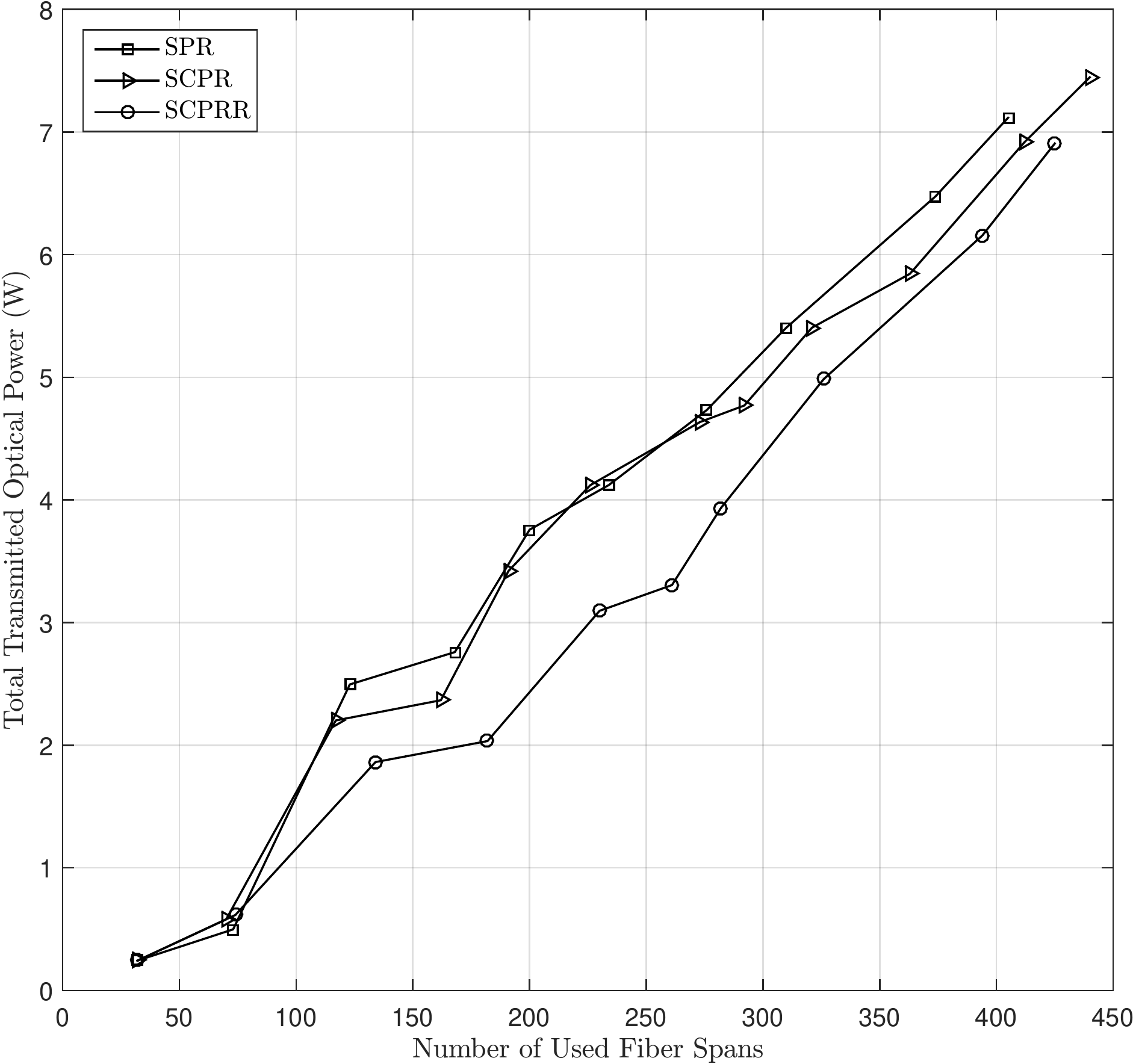}}
\center{\caption{\label{fig:power}  Total transmission optical powers in terms of the number of used fiber spans.}}
\end{figure}

\begin{figure}[t!] 
\center{\includegraphics[scale=0.45]{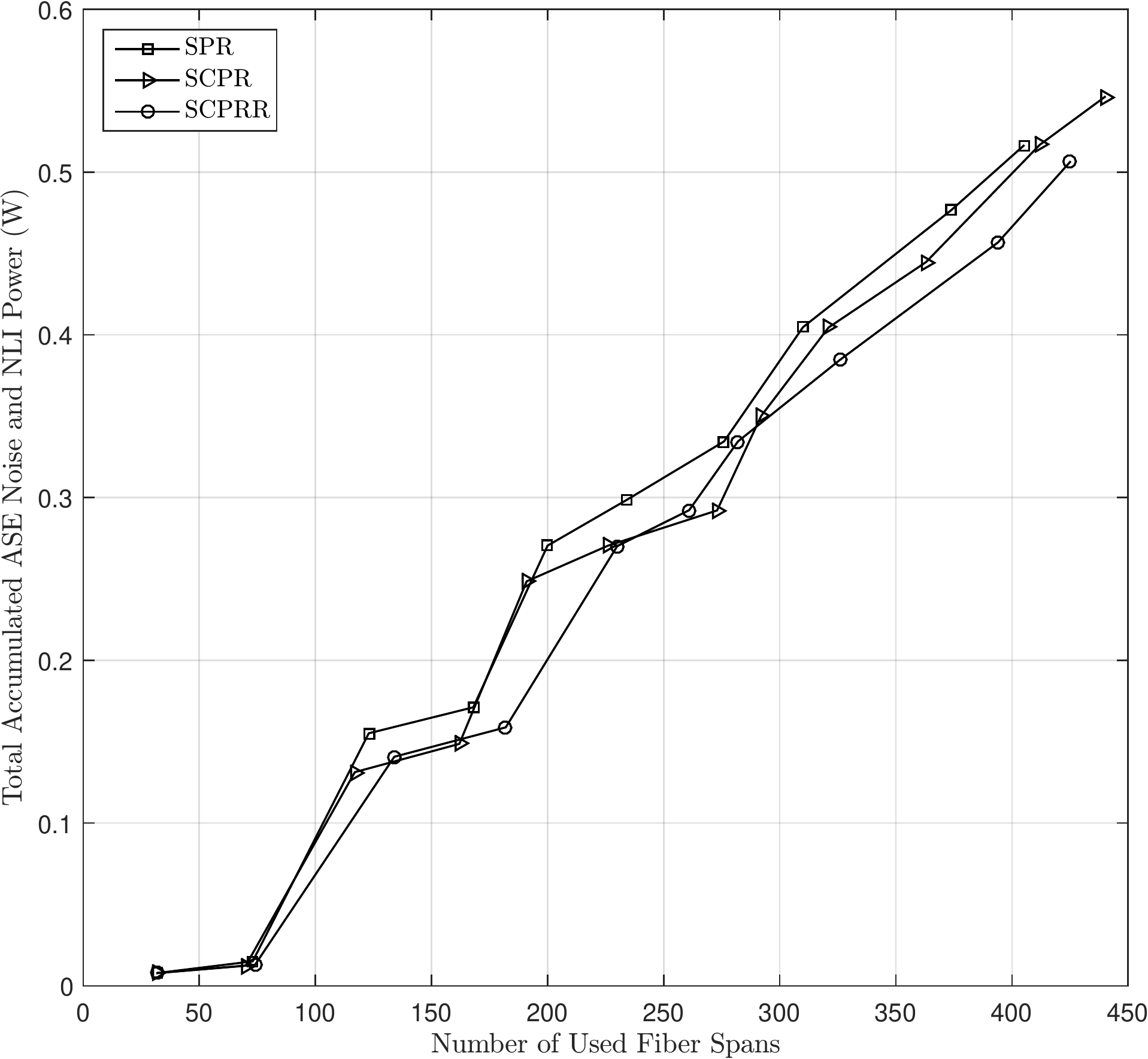}}
\center{\caption{\label{fig:noise}  Total accumulated NLI and ASE noise power in terms of the number of used fiber spans.}}
\end{figure}

Finally, to compare proposed RTO procedures, we run Alg. \ref{alg:rsa} for different RTO algorithms and fixed GPSA1 formulation and compare the total transmitted optical power and accumulated NLI and ASE noise power in terms of the total number of used fiber spans. We use CPLEX \cite{ilog2012cplex} software package for solving BLP and BQP formulations of SPR, SCPR and SCPRR. As shown in Fig. \ref{fig:power} and Fig. \ref{fig:noise}, the amount of transmitted optical power and accumulated NLIs and ASE noise powers are the lowest for SCPRR compared to two other RTO procedures.  The improvement gap is higher for low traffic networks. This shows that SCPRR is more suitable for RTO in EONs with considerable amount of fiber NLIs and ASE noise.
\section{Conclusion}\label{Sec_VII}
In this paper, we provide a heuristic algorithm for routing, traffic ordering, power allocation and spectrum assignment in elastic optical networks in which mixed minimization of transmitted optical power and spectrum usage constrained to physical and quality of service requirements is addressed. Our proposed algorithm consists of two main stages for routing/traffic ordering, and power/spectrum assignment. We propose three routing/traffic ordering procedures and compare them in terms of the total transmission optical power, spontaneous emission noise and nonlinear interferences. We provide posynomial expressions for fiber nonlinear cross- and self-channel interferences and relate modulation spectral efficiency values to their corresponding minimum required optical signal to noise ratio through posynomial curve fitting. Then, we use the results to formulate different geometric optimization problems for power/spectrum assignment stage of the heuristic algorithm.  Although there is a trade-off between run time and accuracy of the proposed geometric formulations, for all of them a considerable reduction in convergence time compared to the well-known mixed integer nonlinear power/spectrum assignment is obtained. Simulation results also confirm that minimization of the number of common fiber spans to transmission bit rate product is the preferred cost function for routing/traffic ordering in elastic optical networks with considerable nonlinear interferences and additive amplification noise. 

\bibliographystyle{IEEEtran}
\bibliography{Reference}

\end{document}